\documentclass[11pt]{article}
\usepackage{amsmath,amssymb,color,graphics,epsfig}

\usepackage{amsfonts}

\newcommand{\be}{\begin{equation}}
\newcommand{\ee}{\end{equation}}
\newcommand{\bea}{\setlength\arraycolsep{2pt} \begin{eqnarray}}
\newcommand{\eea}{\end{eqnarray}}
\newcommand{\nn}{\nonumber}

\def\ft#1#2{{\textstyle{\frac{\scriptstyle #1}{\scriptstyle #2} } }}
\def\fft#1#2{{\frac{#1}{#2}}}

\def\0{{\sst{(0)}}}
\def\1{{\sst{(1)}}}
\def\2{{\sst{(2)}}}
\def\3{{\sst{(3)}}}
\def\4{{\sst{(4)}}}
\def\5{{\sst{(5)}}}
\def\6{{\sst{(6)}}}
\def\7{{\sst{(7)}}}
\def\8{{\sst{(8)}}}
\def\sst#1{{\scriptscriptstyle #1}}

\begin{document}

\begin{flushright}
\end{flushright}

\vspace{25pt}
\begin{center}
{\large {\bf Construction of Regular Black Holes in General Relativity}}

\vspace{10pt}
 Zhong-Ying Fan$^{1}$ and Xiaobao Wang$^{2}$

\vspace{10pt}
{\it $^{1}$Center for High Energy Physics, Peking University, No.5 Yiheyuan Rd, \\}
{\it Beijing 100871, P. R. China\\}
\smallskip
{\it $^{2}$ Department of Physics, Beijing Normal University, Beijing 100875, P. R. China}

\vspace{40pt}

\underline{ABSTRACT}
\end{center}
We present a general procedure for constructing exact black hole solutions with electric or magnetic charges in General Relativity coupled to a nonlinear electrodynamics. We obtain a variety of two-parameter family spherically symmetric black hole solutions. In particular, the singularity at the central of the space-time can be cancelled in the parameters space and the black hole solutions become regular everywhere in the space-time. We study the global properties of the solutions and derive the first law of thermodynamics. We also generalize the procedure to include a cosmological constant and construct regular black hole solutions that are asymptotic to anti-de Sitter space-time.

\vfill {\footnotesize  Email: fanzhy@pku.edu.cn\ \ \ xiaobao@mail.bnu.edu.cn\,.}

\thispagestyle{empty}

\pagebreak

\tableofcontents
\addtocontents{toc}{\protect\setcounter{tocdepth}{2}}




\section{Introduction}
The celebrated singularity theorems proved by Penrose and Hawking \cite{hawking} claim that under some circumstances the existence of singularities is inevitable in General Relativity. This is in accordance with the observation that the first known exact black hole solutions in General Relativity have a singularity inside the event horizon. However, it is widely believed that the singularities are nonphysical objects which are created by classical theories of gravity and they do not exist in nature.
In fact, the quantum arguments given by Sakharov \cite{Sakharov:1966aja} and Gliner \cite{Gliner} suggest that the space-time singularities could be avoided for matter sources with a de Sitter core at the central of the space-time. Based on this idea, Bardeen proposed the first static spherically symmetric regular black hole solution \cite{bardeen}. Other regular black hole models are also proposed later \cite{Borde:1994ai,Barrabes:1995nk,Bogojevic:1998ma,Cabo:1997rm,Hayward:2005gi,Bambi:2013ufa,Ghosh:2014hea,Toshmatov:2014nya,Azreg-Ainou:2014pra,Dymnikova:2015hka}. It is easily shown that all these regular black hole models violate the strong energy condition\footnote{The rotating regular black hole models also violate the weak energy condition.} and hence can break the singularity theorems.

It was established by Ay\'{o}n-Beato and Garc\'{i}a \cite{AyonBeato:1998ub,AyonBeato:1999ec,AyonBeato:1999rg,AyonBeato:2000zs,AyonBeato:2004ih} that the regular black hole models can be interpreted as the gravitational field of a nonlinear
electric or magnetic monopole. Thus, the physical source of the regular black holes could be a nonlinear electromagnetic field. This is also ensured by other authors in the literature \cite{Ma:2015gpa}. Recently, it was shown in \cite{Junior:2015fya} that some regular black hole solutions can be constructed in $f(T)$ gravity coupled to a nonlinear electrodynamics.

In this paper, motivated by the idea of Ay\'{o}n-Beato and Garc\'{i}a we study whether there exists a general procedure for constructing regular black hole solutions in General Relativity coupled to a nonlinear electrodynamics. We find that the answer is yes. In fact, we can construct a lot of static spherically symmetric black hole solutions with two independent integration constants. The regular black holes emerge as some degenerated solutions in the parameters space. We study the thermodynamic properties of the solutions and derive the first law, Smarr formula and entropy product formulae, respectively. We also find that the procedure can be straightforwardly generalized to including a cosmological constant and constructing black hole solutions that are asymptotic to anti-de Sitter space-time.

The paper is organized as follows. In section 2, we study Einstein gravity coupled to a nonlinear electrodynamics and discuss the geometric conditions for regular black holes. In section 3, we construct a variety of magnetically charged black hole solutions in the gravity model. In section 4, we demonstrate the procedure for  constructing electrically charged solutions in the gravity model. In section 5, we study the thermodynamic properties of the solutions above and derive the first law of thermodynamics. In section 6, we generalize the procedure to gravity theories with a cosmological constant and construct AdS black hole solutions.  We conclude this paper in section 7.

\section{Einstein gravity coupled to a non-linear electrodynamics}\label{sec2}
We consider Einstein gravity coupled to a non-linear electromagnetic field of the type
\be I=\fft{1}{16\pi}\int \mathrm{d}^4x\sqrt{-g}\, \Big(R-\mathcal{L}(\mathcal{F})\Big) \,,\label{gravitymodel}\ee
where $F=dA$ is the field strength of the vector field, $\mathcal{F}\equiv F_{\mu\nu}F^{\mu\nu}$ and the Lagrangian density $\mathcal{L}$ is a function of $\mathcal{F}$. The covariant equations of motion are
\be
G_{\mu\nu}=T_{\mu\nu}\,,\qquad \nabla_{\mu}\Big( \mathcal{L}_{\mathcal{F}} F^{\mu\nu}\Big)=0 \,,
\label{eom}\ee
where $G_{\mu\nu}=R_{\mu\nu}-\fft 12 R g_{\mu\nu}$ is the Einstein tensor and $\mathcal{L}_{\mathcal{F}}=\fft{\partial\mathcal{L}}{\partial\mathcal{F}} $. The energy momentum tensor is
\be T_{\mu\nu}=2\Big( \mathcal{L}_{\mathcal{F}} F_{\mu\nu}^2-\fft 14 g_{\mu\nu} \mathcal{L} \Big)  \,.\label{energymomentum}\ee
In this paper, we consider the static spherically symmetric black hole solutions with nonlinear electric/magnetic charges. The most general ansatz is given by
\be ds^2=-f dt^2+\fft{dr^2}{f}+r^2 d\Omega^2 \,,\qquad A=a(r)dt+Q_m \cos{\theta}\, d\phi \,,\label{ansatz}\ee
where $f=f(r)$ and $d\Omega^2=d\theta^2+\sin{\theta}^2 d\phi^2$ denotes the metric of a unit $2$-sphere, $Q_m$ is the total magnetic charge defined by
\be Q_m=\fft{1}{4\pi}\int F \,.\ee
Note that in above ansatz $-g_{tt}=g^{rr}=f$ is consistent with the Einstein equations of motion. This will be shown later in detail.
It turns out that the construction of analytical black hole solutions with dyonic charges is of great difficult. The situation becomes much simpler for single charged case, namely when $a(r)=0$ or $Q_m=0$. Hence, in the following sections we will explicitly show how to construct exact black hole solutions with either magnetic or electric charges.

Since the main motivation of this paper is to construct regular black holes in this gravity model, it is instructive to first discuss what kind of a metric is regular at the origin of the space-time . For this purpose, we parameterize the metric function as
\be f=1-\fft{2m(r)}{r} \,,\label{metricf}\ee
where the constant mass of the Schwarzschild black hole is replaced by a mass distribution function $m(r)$. To govern the existence of an event horizon, we shall require the mass function being positive definite, namely $m(r)>0$ when $r>0$.
To exclude the space-time singularity at the origin, we consider a smooth function $m(r)$ which is at least three times differentiable and approaches zero sufficiently fast in the limit $r\rightarrow 0$: $m(r)\,,m'(r)\,,m''(r)$ vanishes but the third order derivative $m'''(r)$ is finite (zero or nonzero) at the origin $r=0$. Then to ensure the space-time regularity, a sufficient condition is $m(r)/r^3$ is finite in the limit $r\rightarrow 0$ because the curvature polynomials involve at most second order derivatives of the metric. To be concrete, we present some low lying curvature polynomials as follows
\bea
&&R=\fft{4m'}{r^2}+\fft{2m''}{r}\,,\qquad R_{\mu\nu}R^{\mu\nu}=\fft{8m'^2}{r^4}+\fft{2m''^2}{r^2}\,,\nn\\
&& R_{\mu\nu\lambda\rho} R^{\mu\nu\lambda\rho}=\fft{48m^2}{r^6}-\fft{16m}{r^3}\Big(\fft{4m'}{r^2}-\fft{m''}{r} \Big)
+4\Big(\fft{8m'^2}{r^4}-\fft{4m'm''}{r^3}+\fft{m''^2}{r^2} \Big)\,.
\eea
It is clear that if $m(r)/r^3$ is finite and hence $m'(r)/r^2\,,m''(r)/r$ are also finite in the limit $r\rightarrow 0$, all these polynomials will be finite constants at the origin\footnote{Of course, one should further verify that the metric behaves regular everywhere in the space-time when an exact solution satisfying these conditions is successfully constructed.}. Thus, from pure mathematic point of view there exists a variety of candidates for regular black holes in nature except for those with a de Sitter core (namely $m(r)/r^3$ is a finite but nonzero constant) at the central of the space-time. The existence of such regular black hole solutions cannot be ruled out before we have a better understanding of the theory of quantum gravity.

\section{Asymptotically flat black holes with magnetic charges}\label{sec3}
In this section, we will explicitly demonstrate the construction procedure of exact black hole solutions with magnetic charges. In this case, the general ansatz is given by (\ref{ansatz}) with $a(r)=0$. It turns out that the non-linear Maxwell equations are automatically satisfied.
For Einstein equations, we find that there are only two independent equations, given by
\bea
&&0=\fft{f'}{r}+\fft{f-1}{r^2}+\ft 12 \mathcal{L}\,,\\
&&0=f''+\fft{2f'}{r}+\mathcal{L}-\fft{4Q_m^4}{r^4}\mathcal{L}_{\mathcal{F}}\,,
\eea
where a prime denotes the derivative with respect to the radial coordinate. One can first solve the Lagrangian density $\mathcal{L}$ as a function of $r$
\be \mathcal{L}=-2\Big( \fft{f'}{r}+\fft{f-1}{r^2} \Big) \,,\label{la1} \ee
and then substitute it into the second equation. We find that the latter is automatically satisfied for any given metric function $f$. Hence, the metric ansatz (\ref{ansatz}) is indeed most general for static spherically symmetric solutions with magnetic charges.
Under the parametrization (\ref{metricf}), the Lagrangian density simplifies to
\be \mathcal{L}=\fft{4m'(r)}{r^2} \,.\label{la2}\ee
In addition, the square of the field strength $\mathcal{F}$ is
\be \mathcal{F}=\fft{2Q_m^2}{r^4} \,.\label{fs}\ee
Thus, one can freely choose a mass function $m(r)$ which is interesting in physics and then solve the Lagrangian density analytically as a function of $\mathcal{F}$. This completes
the construction of static solutions with magnetic charges. However, there is a potential short coming in this procedure. The magnetic charge $Q_m$ and the integration constants from the metric function $f$ may in general appear in the derived Lagrangian density as well. This means that the solution has no free parameters because all the constants in the solution are the coupling constants of the corresponding theory. As such a solution is less interesting in physics, we shall focus on constructing the solution with at least one free integration constant.

To check the consistency of above procedure, let us discuss two simple examples. The first is when $m(r)=\mathrm{const}$. The metric is a Schwarzschild black hole, which is the solution of vacuum Einstein equations whilst the equation (\ref{la2}) implies $\mathcal{L}=0$, as expected. The second example is
\be m(r)=M-\fft{Q_m^2}{2r} \,.\ee
The metric is a  magnetically charged Reissner-Nordstr\"{o}m black hole, which is the solution of Einstein-Maxwell theories. On the other hand, from the equations (\ref{la2}-\ref{fs}), we find $\mathcal{L}=\mathcal{F}$, as expected.

Using the procedure demonstrated above, we can easily construct a lot of exact black hole solutions with magnetic charges in the gravity model. In the following, we will present three different classes solutions, which include the well-known regular black hole models such as the Bardeen black hole \cite{bardeen} and the Hayward black hole \cite{Hayward:2005gi}.

\subsection{Case 1: Bardeen class}
The first class solution that we present is valid for a Lagrangian density
\be \mathcal{L}=\fft{4\mu}{\alpha}\fft{\big(\alpha \mathcal{F} \big)^{5/4}}{\big( 1+\sqrt{\alpha \mathcal{F}} \big)^{1+\mu/2}} \,,\label{lagrangian1}\ee
where $\mu>0$ is a dimensionless constant and $\alpha>0$ has the dimension of length squared. In the weak filed limit, the vector field behaves as $\mathcal{L}\sim \alpha^{1/4}\mathcal{F}^{5/4}$ which is slightly stronger than a Maxwell filed. The general two-parameter family black hole solution is
\bea\label{sol1}
&&ds^2=-f dt^2+\fft{dr^2}{f}+r^2 d\Omega^2 \,,\qquad A=Q_m \cos{\theta}\, d\phi \,,\nn\\
&&f=1-\fft{2M}{r}-\fft{2\alpha^{-1} q^3 r^{\mu-1}}{\big( r^2+q^2 \big)^{\mu/2}}\,,
\eea
where $q$ is a free integration constant which is related to the magnetic charge
\be Q_m=\fft{q^2}{\sqrt{2\alpha}} \,.\label{magnetic}\ee
For the same physical charge, the parameter $q$ can be either positive or negative. However, for our purpose we shall require $q>0$ (through out this paper) because the solution with $M=0$ will no longer be a black hole in the $q<0$ case\footnote{For some of the solutions in this section such as (\ref{sol3}), there exists an additional singularity at $r=-q>0$ when $q<0$, which can be covered by an event horizon even for $M=0$. However, in this case the graviton mode becomes ghost-like. Thus we always require $q>0$ in this paper.}. Note that the solution reduces to a Schwarzschild black hole in the neutral limit. For $M=0\,,\mu=3$, the solution is the Bardeen black hole \cite{bardeen}, which was first constructed in \cite{AyonBeato:2000zs}. For later convenience, we refer to the parameter $M$ as {\it Schwarzschild mass}. The ADM mass of the black hole can be read off from the asymptotic behavior of the metric functions
\be f=1-\fft{2\big(M+\alpha^{-1}q^3\big)}{r}+\cdots \,.\ee
We have
\be M_{\mathrm{ADM}}=M+M_{\mathrm{em}}\,,\qquad M_{\mathrm{em}}=\alpha^{-1}q^3 \,.\label{adm}\ee
It is worth pointing out that the ADM mass has two copies of contributions, one is the {\it Schwarzschild mass} which describes the condensate of the massless graviton from its nonlinear self-interactions and the other is a charged term which is associated with the nonlinear interactions between the graviton and the photon. The latter contribution is impossible for a Maxwell field or a Born-Infeld field.
\begin{figure}[ht]
  \centering
  \includegraphics[width=230pt]{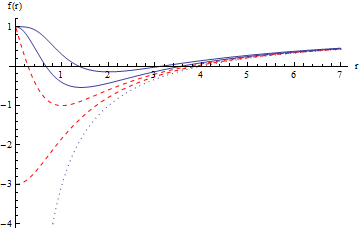}
  \caption{\it The metric function $f(r)$ for Bardeen class solution with zero {\it Schwarzschild mass}. Along the vertical axis, the $\mu$'s value decreases from top to bottom. For solid lines $\mu=5\,,3$ and for dashed lines $\mu=2\,,1$. Some parameters have been set as $\alpha=1/2\,,q=1$. The dotted line corresponds to a Schwarzschild black hole with $M=2$. }\label{f1}
\end{figure}
Since the $M$ term introduces an unavoidable space-time singularity, we focus on discussing the degenerate case with zero {\it Schwarzschild mass}. The metric function $f(r)$ for various $\mu$ is depicted in Fig. \ref{f1}. It is clear that for $\mu\geq 1$, $f(r)$ approaches a finite constant in the limit $r\rightarrow 0$. In fact, near the origin, the metric function behaves as
\be f=1-2\alpha^{-1}q^{3-\mu}r^{\mu-1}+\cdots \,.\label{forigion}\ee
As emphasized earlier, to exclude the space-time singularity the mass function of the solution should satisfy the condition $m(r)/r^3\sim$ cons in the limit $r\rightarrow 0$. This selects a special class solution which has $\mu\geq 3$. Calculating the low lying curvature polynomials, we find
\bea
&&R=\mathrm{regular\,\, term}\times r^{\mu-3}\,,\qquad R_{\mu\nu}R^{\mu\nu}=\mathrm{regular\,\, term}\times r^{2\mu-6}\,,\nn\\
&& R_{\mu\nu\lambda\rho} R^{\mu\nu\lambda\rho}=\mathrm{regular\,\, term}\times r^{2\mu-6}\,,
\label{curva}\eea
where ``regular term" denotes the terms that have a regular limit at the origin. Therefore, for $\mu\geq 3$ the singularity at the origin is indeed cancelled. Finally, we remark that for generic $\mu$, the solution violates the strong energy condition while the weak energy condition is still preserved.

\subsection{Case 2: Hayward class}
The second class solution we present is valid for a Lagrangian density
\be \mathcal{L}=\fft{4\mu}{\alpha}\fft{\big(\alpha \mathcal{F} \big)^{\fft{\mu+3}{4}}}
{\Big(1+\big(\alpha \mathcal{F} \big)^{\fft{\mu}{4}} \Big)^2} \,.\label{lagrangian2}\ee
In the weak field limit, the vector field behaves as $\mathcal{L}\sim \alpha^{\fft{\mu-1}{4}}\mathcal{F}^{\fft{\mu+3}{4}}$. It could be either stronger ($\mu>1$) or weaker ($0<\mu<1$) than a Maxwell field. A critical case occurs when $\mu=1$ at which the nonlinear electrodynamics reduces to a Maxwell field in the weak field limit. The general static spherically symmetric solution reads
\bea\label{sol2}
&&ds^2=-f dt^2+\fft{dr^2}{f}+r^2 d\Omega^2 \,,\qquad A=Q_m \cos{\theta}\, d\phi \,,\nn\\
&&f=1-\fft{2M}{r}-\fft{2\alpha^{-1} q^3 r^{\mu-1}}{ r^\mu+q^\mu}\,,
\eea
where the magnetic charge and the ADM mass are still given by (\ref{magnetic}) and (\ref{adm}) respectively. For $M=0\,,\mu=3$, the solution is the Hayward black hole \cite{Hayward:2005gi} which has been constructed in \cite{Fan:2016rih}. For the solution with zero {\it Schwarzschild mass}, the behavior of the metric function $f$ and the low lying curvature polynomials are still given by (\ref{forigion}) and (\ref{curva}) respectively. Thus, the regular black hole solution has $\mu\geq 3$ as well.

\subsection{Case 3: a new class}
Perhaps the most interesting theories which admit regular black hole solutions are such that the vector field approaches a Maxwell field in the weak field limit. We find that such theories indeed exist
\be \mathcal{L}=\fft{4\mu}{\alpha}\fft{\alpha \mathcal{F}}
{\Big(1+\big(\alpha \mathcal{F} \big)^{1/4} \Big)^{\mu+1}} \,.\label{lagrangian3}\ee
The black hole solution reads
\bea\label{sol3}
&&ds^2=-f dt^2+\fft{dr^2}{f}+r^2 d\Omega^2 \,,\qquad A=Q_m \cos{\theta}\, d\phi \,,\nn\\
&&f=1-\fft{2M}{r}-\fft{2\alpha^{-1} q^3 r^{\mu-1}}{ \big(r+q\big)^\mu}\,,
\eea
where the magnetic charge and ADM mass are still given by (\ref{magnetic}) and (\ref{adm}) respectively. When $M=0$, the metric function $f$ and the curvature polynomials behave as (\ref{forigion}) and (\ref{curva}) near the origin. Thus, the regular black hole solution also has $\mu\geq 3$.

\subsection{Generic case}
For a generic mass function $m(r)$ of the type
\be m(r)=M+\fft{\alpha^{-1}q^3 r^\mu}{\big(r^\nu+q^\nu\big)^{\mu/\nu}} \,.\label{massfunction}\ee
the Lagrangian density of the nonlinear electromagnetic field turns out to be
\be \mathcal{L}=\fft{4\mu}{\alpha}\fft{\big(\alpha \mathcal{F} \big)^{\fft{\nu+3}{4}}}
{\Big(1+\big(\alpha \mathcal{F} \big)^{\fft{\nu}{4}} \Big)^\fft{\mu+\nu}{\nu}} \,,\label{lagrangiangene}\ee
where an extra dimensionless parameter $\nu$ is introduced. The Lagrangian density reduces to (\ref{lagrangian1}), (\ref{lagrangian2}) and (\ref{lagrangian3}) when $\nu=2\,,\mu\,,1$, respectively. For later convenience, we also write down the corresponding black hole solution as follows
\bea\label{solgene}
&&ds^2=-f dt^2+\fft{dr^2}{f}+r^2 d\Omega^2 \,,\qquad A=Q_m \cos{\theta}\, d\phi \,,\nn\\
&&f=1-\fft{2M}{r}-\fft{2\alpha^{-1} q^3 r^{\mu-1}}{ \big(r^\nu+q^\nu\big)^{\mu/\nu}}\,.
\eea
Following the analysis above, the regular black hole solution has $M=0$ and $\mu \geq 3$. Of course, one can also consider to construct black hole solutions with more general mass functions (For example, the first exact regular black hole solutions presented by Ay\'{o}n-Beato and Garc\'{i}a in \cite{AyonBeato:1998ub}. In this paper, we only consider the black hole solutions with the mass function (\ref{massfunction}) because the solutions can have one independent integration constant. For more general mass functions, this condition will no longer
hold but one can still construct those solutions using our procedure) if the difficulty of solving the Lagrangian density analytically can be overcomed.

\section{Asymptotically flat black holes with electric charges}
 For electrically charged black hole solutions, the general ansatz is given by (\ref{ansatz}) with $Q_m=0$. In this case, we find that there are three independent equations
\bea
&&0=\fft{a''}{a'}+\fft{2}{r}+\fft{\mathcal{L}_\mathcal{F}'}{\mathcal{L}_\mathcal{F}}\,,\nn\\
&&0=f''+\fft{2f'}{r}+\mathcal{L}\,,\nn\\
&&0=\fft{f'}{r}+\fft{f-1}{r^2}+\ft 12 \mathcal{L}+2a'^2\mathcal{L}_\mathcal{F}\,,
\eea
where the first is the equation of the vector field which can be solved as
\be \mathcal{L}_\mathcal{F}=\fft{Q_e}{r^2a'}\,.\label{elec1}\ee
Here $Q_e$ is the electric charge carried by the black hole
\be Q_e=\fft{1}{4\pi}\int \mathcal{L}_{\mathcal{F}} {}^*F\,.\ee
The Lagrangian density can be solved from the second equation
\be \mathcal{L}=-f''-\fft{2f'}{r} \,.\label{elec2}\ee
Substituting above results into the last equation, one finds
\be 0=f''-\fft{2(f-1)}{r^2}-\fft{4Q_e a'}{r^2} \,.\label{elec3}\ee
This is the equation that one should solve to obtain the electric field for a given metric function. To check the consistency of the procedure, we calculate
$\mathcal{L}_\mathcal{F}$ from its definition $\mathcal{L}_\mathcal{F}=\partial\mathcal{L}/\partial\mathcal{F}=\mathcal{L}'/\mathcal{F}'\,,\mathcal{F}=-2a'^2$ and use the equations (\ref{elec2}) and (\ref{elec3}). We find that the result exactly coincides with (\ref{elec1}).

Under the parametrization (\ref{metricf}), the Lagrangian density simplifies to
\be \mathcal{L}=\fft{2m''}{r} \,.\ee
The equation (\ref{elec3}) can be analytically solved as
\be a=\ft{1}{2Q_e}\big( 3m-r m' \big)+c \,,\ee
where $c$ is an integration constant associated with the gauge choice. In the following, we shall choose the gauge $a(\infty)=0$. This completes the construction of black hole solutions with nonlinear electric charges. One can first choose a physically interesting mass function and then obain the corresponding gauge potential and the Lagrangian density as a function of $r$ from above two equations. The remaining problem is how to express the Lagrangian density explicitly as a function of the field strength squared $\mathcal{F}$. In general, this is very difficult because $\mathcal{F}=-2a'^2$ has a rather complicated expression\footnote{The situation becomes much simpler for a Maxwell and Born-Infeld field.}. Nevertheless, the equation (\ref{elec1}) allows us to rewrite the Lagrangian density at least as a function of $P$ where $P=\mathcal{F}\big(\mathcal{L}_\mathcal{F}\big)^2$, namely $\mathcal{L}=\mathcal{L}(P)$. In fact, in this case it may be more appropriate to describe the system by meas of a Legendre transformation \cite{AyonBeato:1998ub,Salazar:1987ap}
\be \mathcal{H}=\mathcal{F}\mathcal{L}_{\mathcal{F}}-\mathcal{L} \,.\ee
It is easily to show that $\mathcal{H}$ is naturally a function of $P$: $d\mathcal{H}=\big(\mathcal{L}_\mathcal{F}\big)^{-1}d\Big(\mathcal{F}\big(\mathcal{L}_{\mathcal{F}}\big)^2 \Big)=\mathcal{H}_P dP$ and the Lagrangian density can be derived as $\mathcal{L}=2P\mathcal{H}_P-\mathcal{H}$. It should be emphasized that the original $\mathcal{L}(\mathcal{F})$ formalism may not be appropriate any longer in this case\footnote{We are grateful to E.~Ayon-Beato for this point.} because one will end with a multi-valued $\mathcal{L}(\mathcal{F})$, which have different branches for a well-defined single one $\mathcal{H}(P)$.

For the generic mass function (\ref{massfunction}), the electrically charged black hole solution reads
\bea\label{solgene2}
&&ds^2=-f dt^2+\fft{dr^2}{f}+r^2 d\Omega^2 \,,\qquad f=1-\fft{2M}{r}-\fft{2\alpha^{-1} q^3 r^{\mu-1}}{ \big(r^\nu+q^\nu\big)^{\mu/\nu}}\,,\nn\\ &&A=\fft{q}{\sqrt{2\alpha}}\Big(\big(3-(\mu-3)(\ft{q}{r})^\nu\big)\big(1+(\ft{q}{r})^\nu\big)^{-\fft{\mu+\nu}{\nu}}-3\Big) \, dt \,,
\eea
where the electric charge is given by
\be Q_e=\fft{q^2}{\sqrt{2\alpha}} \,.\ee
The corresponding Lagrangian density can be solved as
\be \mathcal{L}=\fft{2\mu}{\alpha}z^{\mu-3}\big(1+z^\nu\big)^{-\fft{\mu+2\nu}{\nu}}\Big(\mu-1-(\nu+1)z^\nu\Big)\,,\qquad z=\fft{q}{r}=\ft{1}{\big(-\alpha P\big)^{1/4}} \,.\label{lagrangiangene2}\ee
Unlike the magnetically charged case,
the field strength of the nonlinear electromagnetic filed can behave regular for some cases. We find
\bea
\mathcal{F}&=&-2a'^2=-\ft{1}{2Q_e^2}\big(2m'-r m''\big)^2 \,,\nn\\
&=&-\alpha^{-1}\mu^2q^{2\nu+2}r^{2\mu-2}\Big((\nu+3)r^\nu-(\mu-3)q^\nu \Big)^2\big(r^\nu+q^\nu\big)^{-\fft{2\mu+4\nu}{\nu}}\,.
\eea
It is easy to see that when $\mu\geq 1$, the electric field has a regular limit at the origin of space-time. Thus, in the subset of the parameters space $M=0\,,\mu\geq 3$ we have regular black holes with regular electric fields.

To end this section, we point out that the construction will become much more complicated for dyonic regular black hole solutions (\ref{ansatz}) because the field strength squared becomes $\mathcal{F}=\fft{2Q_m^2}{r^4}-2 a'^2$ whilst $\mathcal{L}_\mathcal{F}$ still takes the form (\ref{elec1}). Thus, in this case it is of great difficult to solve the Lagrangian density $\mathcal{L}$ as a function of $\mathcal{F}$ or $P$ for a given mass function. Perhaps one can start the story from a given Lagrangian density such as (\ref{lagrangiangene}) and then try to solve the electric filed as well as the metric functions analytically or half analytically.
We leave this as a future direction for research.

\section{The first law of thermodynamics}

\subsection{Derivation of the first law}
For asymptotically flat black holes with nonlinear electric/magnetic charges, the first law was derived in \cite{Rasheed:1997ns} using a covariant approach.  It was shown that the standard first law
\be dM_{\mathrm{ADM}}=T dS+\Phi dQ_e+\Psi dQ_m \,,\ee
was satisfied. Here $T\,,S$ are the Hawking temperature and entropy
\be T=\fft{\kappa}{2\pi}\,,\qquad S=\fft 14 A \,,\ee
where $\kappa$ is the surface gravity and $A$ is the area of the event horizon. The physical charges $Q_e\,,Q_m$ and the conjugate potentials $\Phi\,,\Psi$ are defined by
\bea
&&Q_e=\fft{1}{4\pi}\int_{\Sigma_2}\mathcal{L}_{\mathcal{F}} {}^*F\,,\qquad \Phi=A_t(\infty)-A_t(r_0) \,,\nn\\
&&Q_m=\fft{1}{4\pi}\int_{\Sigma_2} F \,,\qquad  \Psi=\widetilde{A}_t(r_0)-\widetilde{A}_t(\infty)\,,\qquad \widetilde{F}=d\widetilde{A}=\mathcal{L}_{\mathcal{F}}\, {}^*F \,.
\label{definition}\eea
Note that the definitions for the electric charge and magnetic potential are properly generalized\footnote{To benefit the readers, we shall briefly explain how we arrive at the definitions in (\ref{definition}) for a nonlinear electrodynamics. As usual the equations of motion and the Bianchi identities can be expressed as $ d\widetilde{F}=0\,,dF=0$.
Then the physical charges can be defined by integrating the l.h.s of the equations over any closed-2 surface enclosing the charges. The electric and magnetic field vectors can be defined by $E_\mu=F_{\mu\nu}\xi^\nu\,, B_\mu=-\widetilde{F}_{\mu\nu}\xi^\nu$, where $\xi$ is a Killing vector that is null on the black hole event horizon (here our discussions are valid for generally stationary and axi-symmetric black hole solutions). Using the equations of motion and Bianchi identities, one can show that $\nabla_{[\mu}E_{\nu]}=0=\nabla_{[\mu}B_{\nu]}$ due to time-translational and rotational symmetries. Hence the electric/magnetic field vectors can be written as $E_\mu=\partial_\mu \Phi\,,B_\mu=\partial_\mu \Psi$, which in fact defines the electric/magnetic potentials covariantly. Moreover, it is easily shown that
$\Phi=A_\mu \xi^\mu$ and $\Psi=-\widetilde{A}_\mu \xi^\mu $, up to a gauge choice. In static space-times, this gives the definitions in (\ref{definition}).} and they coincide with the conventional one for a linear Maxwell field (for more details, we suggest the interested readers referring to \cite{Rasheed:1997ns}).
Furthermore, if $\alpha$ is taken as a thermodynamic variable, the first law generalized in the extended phase space reads\footnote{It is shown by Zhang and Gao \cite{Zhang:2016ilt} that the additional terms in the first law for asymptotically flat black holes can be derived using the covariant approach in \cite{Rasheed:1997ns}.}
\be dM_{\mathrm{ADM}}=T dS+\Phi dQ_e+\Psi dQ_m+\Pi d\alpha \,,\label{firstlaw}\ee
where $\Pi$ is a new quantity conjugate to $\alpha$. It is defined by
\be \Pi=\fft 14\int_{r_0}^{\infty}\mathrm{d}r\,\sqrt{-g}\,\fft{\partial\mathcal{L}}{\partial\alpha} \,.\ee
Note that $\Pi$ has the dimension of energy. Then the scaling dimensional argument\footnote{Euler's theorem implies that for any given function $g(x_i)$ such that $\mu^\delta g(x_i)=g(\mu^{\delta_i} x_i)$, one has $ \delta g(x_i)=\delta_i x_i \fft{\partial g}{\partial x_i} $. Here $\delta\,,\delta_i $ denote the scaling dimensions of the function $g(x_i)$ and the variables $x_i$ respectively. For our first law (\ref{firstlaw}), we have $[M_{\mathrm{ADM}}]=L\,,[S]=L^2\,,[Q_e]=L\,,[Q_m]=L\,,[\alpha]=L^2$, implying that the Smarr formula is (\ref{smarr}).} implies that the Smarr formula is
\be M_{\mathrm{ADM}}=2TS+\Phi Q_e+\Psi Q_m+2 \Pi \alpha \,.\label{smarr}\ee
It is worth pointing out that the existence of the new conjugate ($\Pi\,,\alpha$) is essential to govern the Smarr formula. However, the definition of the conjugate is not unique. One can redefine a new quantity $\widetilde{\alpha}\sim \alpha^z$ and its conjugate as $\widetilde{\Pi}d\widetilde{\alpha}=\Pi d\alpha$. Then the Smarr formula (\ref{smarr}) holds with the term $2\Pi \alpha$ replaced by $2z\widetilde{\Pi}\widetilde{\alpha}$. An interesting question is
the physical interpretation of the new pair of conjugate which however remains open and deserves further studies.

In the following, we shall test the first law (\ref{firstlaw}) and the Smarr formula (\ref{smarr}) for the exact solutions that we construct previously.
First, for the magnetically charged solutions (\ref{solgene}), the various thermodynamic quantities are given by
\bea
&&M_{\mathrm{ADM}}=M+\alpha^{-1}q^3 \,,\nn\\
&&S=\pi r_0^2\,,\quad T=\fft{1}{4\pi r_0}\Big(1-2\mu\alpha^{-1}q^4 r_0^{\mu-1}(r_0+q)^{-\mu-1} \Big) \,,\nn\\
&&Q_m=\fft{q^2}{\sqrt{2\alpha}}\,,\qquad \Psi=-\fft{q}{\sqrt{2\alpha}}\Big(\big(3-(\mu-3)\ft{q}{r_0}\big)\big(1+\ft{q}{r_0}\big)^{-\mu-1}-3\Big)\,,\nn\\
&&\Pi=\fft{q^3}{4\alpha^2}\Big( \big(1+(\mu+1)\ft{q}{r_0} \big)\big(1+\ft{q}{r_0}\big)^{-\mu-1} -1\Big)\,.
\label{thermquantitiy1}\eea
It follows that the first law (\ref{firstlaw}) and the Smarr formula (\ref{smarr}) with vanishing electric charge hold straightforwardly.

For the electrically charged solutions (\ref{solgene2}), we have
\bea
&&M_{\mathrm{ADM}}=M+\alpha^{-1}q^3 \,,\nn\\
&&S=\pi r_0^2\,,\quad T=\fft{1}{4\pi r_0}\Big(1-2\mu\alpha^{-1}q^{\nu+3} r_0^{\mu-1}\big(r_0^\nu+q^\nu\big)^{-\fft{\mu+\nu}{\nu}} \Big) \,,\nn\\
&&Q_e=\fft{q^2}{\sqrt{2\alpha}}\,,\qquad \Phi=-\fft{q}{\sqrt{2\alpha}}\Big(\big(3-(\mu-3)(\ft{q}{r_0})^\nu\big)\big(1+(\ft{q}{r_0})^\nu\big)^{-\fft{\mu+\nu}{\nu}}-3\Big)\,,\nn\\
&&\Pi=\fft{q^3}{4\alpha^2}\Big( \big(1+(\mu+1)(\ft{q}{r_0})^\nu \big)\big(1+(\ft{q}{r_0})^\nu\big)^{-\fft{\mu+\nu}{\nu}} -1\Big)\,.
\label{thermquantitiy2}\eea
It is straightforward to verify that the first law (\ref{firstlaw}) and the Smarr formula (\ref{smarr}) with vanishing magnetic charge are indeed satisfied.
\subsection{Entropy product formulae}
Let us discuss the entropy product formulae for the solutions (\ref{solgene}) and (\ref{solgene2}).
For simplicity, we focus on the three special classes solutions listed in section \ref{sec3} with some low lying $\mu=1\,,2\,,3$. For vanishing {\it Schwarzschild mass}, the maximal number of the horizons defined by the roots (both real and imaginary) of the equation $f(r)=0$ is exactly equal to $\mu$ for all these solutions. For instance, for Bardeen class solutions (\ref{sol1}), there is only one horizon $r_0=\sqrt{4M_{\mathrm{em}}^2-q^2}$ for $\mu=1$ and two horizons $r_\pm=M_{\mathrm{em}}\pm \sqrt{M_{\mathrm{em}}^2-q^2}$ for $\mu=2$. In both cases, the reality of the horizons provides a lower bound for the physical charges: $Q>\mu \sqrt{\alpha/8}$. For $\mu=3$, the equation $f(r)=0$ is equivalent to a cubic equation of $r^2$
\be 0=(r^2+q^2)^3-4M_{\mathrm{em}}^2r^4 \,,\ee
and hence there are six roots in total, which occurs in pairs with $r^2$ taking the same value. Here we follow \cite{Cvetic:2010mn} and view $\tilde{r}=r^2$ as the radial variable and consider only three roots. We find that the horizon radius have lengthy expressions which are not instructive to give. Nevertheless, the entropy product formula turns out to be very simple. We find

\bea
&&\mu=1\,,\qquad S=\pi \Big(4M_{\mathrm{em}}^2-\sqrt{2\alpha}\,Q \Big)\,,\nn\\
&&\mu=2\,,\qquad \prod_{i=1}^{2}S_i=2\alpha \pi^2 Q^2\,,\nn\\
&&\mu=3\,,\qquad \prod_{i=1}^{3}S_i=-(2\alpha)^{3/2} \pi^3 Q^3\,,
\eea
where $Q$ collectively denotes the electric/magnetic charges. For $\mu=4$, we can also derive the product $\prod_{i=1}^4 S_i$ which is a rather involved function of ($M_{\mathrm{em}}\,,\alpha^{1/2}Q$). For Hayward class (\ref{sol2}) and the new class solutions (\ref{sol3}), we obtain the same entropy product formulas
\bea
&&\mu=1\,,\qquad S=\pi \Big(2M_{\mathrm{em}}^2-(2\alpha)^{1/4}\,Q^{1/2}\Big)^2\,,\nn\\
&&\mu=2\,,\qquad \prod_{i=1}^{2}S_i=2\alpha \pi^2 Q^2\,,\nn\\
&&\mu=3\,,\qquad \prod_{i=1}^{3}S_i=(2\alpha)^{3/2} \pi^3 Q^3\,.
\eea
Note that for all the solutions above, the entropy product formulas for $\mu=2\,,3$ are independent of $M_{\mathrm{em}}$.

For the solutions with nonzero {\it Schwarzschild mass}, we can also derive the product formulae of the entropies for $\mu=1\,,2$ cases. We find
\bea
&&\mu=1\,,\qquad \prod_{i=1}^2S_i=4(2\alpha)^{1/2}\pi^2 M^2 Q\,,\nn\\
&&\mu=2\,,\qquad \prod_{i=1}^{3}S_i=8\alpha \pi^3 M^2 Q^2\,,
\eea
which intriguingly depend on the product of the {\it Schwarzschild mass} squared and the physical charges. Note that the $\mu=1$ case of the Bardeen class solution has four horizons and hence is not included in above results. The entropy product of this case as well as the $\mu=3$ case of all these solutions is in general a rather involved function of $(M\,,M_{\mathrm{em}}\,,\alpha^{1/2}Q)$.
\section{Asymptotically anti-de Sitter black holes}
The charged AdS black holes play an important role in the application of the AdS/CFT correspondence. In this section, we would like to construct the AdS black hole solutions with nonlinear electric/magnetic charges. For this purpose, we include a cosmological constant in the action, namely
\be I=\fft{1}{16\pi}\int \mathrm{d}^4x\sqrt{-g}\, \Big(R+6\ell^{-2}-\mathcal{L}(\mathcal{F})\Big) \,,\ee
where $\ell$ is the AdS radius. The covariant equations of motion are still given by (\ref{eom}-\ref{energymomentum}) but the Einstein tensor includes the
cosmological constant $G_{\mu\nu}=R_{\mu\nu}-\fft 12 (R+6\ell^{-2}) g_{\mu\nu}$. We find that for maximally symmetric solutions with electric/magnetic charges, the procedure established in section \ref{sec2} and \ref{sec3} still works well (Of course, some of the equations involve new terms associated with the cosmological constant). Here we shall not repeat those details. The final results are for the same Lagrangian density $\mathcal{L}(\mathcal{F})$, the asymptotically flat black hole solutions obtained in section \ref{sec2} and \ref{sec3} can be straightforwardly generalized to (A)dS black hole solutions with spherical/hyperbolic/toric topologies. For the magnetic case (\ref{lagrangiangene}), the solution reads
\bea\label{solgene3}
&&ds^2=-f dt^2+\fft{dr^2}{f}+r^2 d\Omega_{k}^2 \,,\qquad A=Q_m \,x dy \,,\nn\\
&&f=r^2/\ell^2+k-\fft{2M}{r}-\fft{2\alpha^{-1} q^3 r^{\mu-1}}{ \big(r^\nu+q^\nu\big)^{\mu/\nu}}\,,
\eea
where $d\Omega_k^2=dx^2/(1-k x^2)+(1-k x^2)dy^2$ denotes the metric of the two dimensional sphere/hyperboloid/torus with constant curvature $k=1\,,-1
\,,0$.

For the electrically charged case (\ref{lagrangiangene2}), the solution reads
\bea\label{solgene4}
&&ds^2=-f dt^2+\fft{dr^2}{f}+r^2 d\Omega_k^2 \,,\qquad f=r^2/\ell^2+k-\fft{2M}{r}-\fft{2\alpha^{-1} q^3 r^{\mu-1}}{ \big(r^\nu+q^\nu\big)^{\mu/\nu}}\,,\nn\\ &&A=\fft{q}{\sqrt{2\alpha}}\Big(\big(3-(\mu-3)(\ft{q}{r})^\nu\big)\big(1+(\ft{q}{r})^\nu\big)^{-\fft{\mu+\nu}{\nu}}-3\Big) \, dt \,,
\eea
Treating the cosmological constant as well as the parameter $\alpha$ as a thermodynamic variable \cite{Kastor:2009wy,Cvetic:2010jb}, we argue that the first law in the extended phase space reads
\be dM_{\mathrm{AMD}}=T dS+\Phi dQ_e+\Psi dQ_m+\Pi d\alpha+V d\widetilde{P} \,,\label{firstlaw2}\ee
where $M_{\mathrm{AMD}}$ is the AMD mass \cite{Ashtekar:1984zz,Ashtekar:1999jx} of AdS black holes and the conjugate $(\widetilde{P}\,,V)$ are defined by \cite{Kastor:2009wy,Cvetic:2010jb}
\be \widetilde{P}=-\fft{\Lambda}{8\pi}=\fft{3}{8\pi\ell^2}\,,\qquad V=\fft{4\pi r_0^3}{3} \,.\ee
The Smarr formula is
\be M_{\mathrm{ADM}}=2TS+\Phi Q_e+\Psi Q_m+2 \Pi \alpha-2V \widetilde{P} \,.\label{smarr2}\ee
To test the first law and Smarr formula for above solutions, we first notice that the temperature of the solutions has additional dependence on the cosmological constant as well as the topological parameter. We find
\bea
&&\mathrm{magnetic\,\, solution}:\quad T=\fft{1}{4\pi r_0}\Big(3r_0^2\ell^{-2}+k-2\mu\alpha^{-1}q^4 r_0^{\mu-1}(r_0+q)^{-\mu-1} \Big)\,,\nn\\
&&\mathrm{electric\,\,solution}:\quad T=\fft{1}{4\pi r_0}\Big(3r_0^2\ell^{-2}+k-2\mu\alpha^{-1}q^{\nu+3} r_0^{\mu-1}\big(r_0^\nu+q^\nu\big)^{-\fft{\mu+\nu}{\nu}} \Big)\,.
\label{tem}\eea
The mass and other thermodynamic quantities exactly coincide with those of (\ref{thermquantitiy1}) and (\ref{thermquantitiy2}) respectively. It follows that the above first law and Smarr formula are indeed satisfied for these solutions.

To end this section, we point out that the equation (\ref{tem}) in fact gives the equation of state $\widetilde{P}=\widetilde{P}(T\,,V)$ for the black hole systems in the extended phase space. Then one can follow \cite{Fan:2016rih} and discuss the critical phenomena of these solutions.

\section{Conclusion}
In this paper, we provide a generic procedure for constructing exact black hole solutions with electric or magnetic charges in General Relativity coupled to a nonlinear electrodynamics. The Lagrangian density of the nonlinear electromagnetic field is proposed to be a function of the field strength squared $\mathcal{F}=F_{\mu\nu}F^{\mu\nu}$. For general static spherically symmetric solutions, we find that the equations of motion allows us to choose an appropriate
metric and then solve the gauge potential and the Lagrangian density of the nonlinear electromagnetic field. This is a simple but powerful procedure for constructing known regular black hole models in the literature.

We first construct magnetically charged solutions in the gravity model. We obtain a large class solutions and derive the corresponding Lagrangian density of the nonlinear electromagnetic field analytically. The black hole solutions contain two free parameters and reduce to the Schwarzschild black hole in the neutral limit. In particular, in a subset of the parameters space the singularity at the origin of the space-time is cancelled and the black holes become regular everywhere in the space-time. We also establish the procedure for constructing electrically charged solutions in the gravity model. We find that in this case, all the regular black holes have a regular electric field as well.

We then study the global properties of above solutions. We derive the first law and the Smarr formula. For some of the solutions, we also derive the entropy product formulae and obtain many interesting results.

Finally, we generalize the construction procedure for gravity theories with a cosmological constant. We find that the above asymptotically flat black hole solutions (including the regular black holes) can be straightforwardly generalized to the maximally symmetric counterparts that are asymptotic to anti-de Sitter space-time.

\section*{Acknowledgments}
 We are grateful to Sijie Gao and E.~Ayon-Beato for valuable correspondence and discussions. Z. Y. Fan is supported in part by NSFC Grants No.~11275010, No.~11335012 and No.~11325522. X. Wang is supported in part by NSFC Grants No. 11235003, No. 11375026 and NCET-12-0054.

\end{document}